\documentclass[11pt, twoside]{article}

\usepackage{amssymb}
\usepackage{amsmath}
\usepackage{array}
\usepackage{arydshln}
\usepackage[dvipsnames]{xcolor}
\usepackage{multirow}
\usepackage{pifont}
\usepackage[margin=1.1in]{geometry}
\usepackage{tabto}
\usepackage{array}
\usepackage{setspace}
\usepackage{listings}
\usepackage{enumitem}
\usepackage[colorlinks=true, citecolor=blue]{hyperref}
\usepackage{tikz}
\usetikzlibrary{arrows, decorations.pathmorphing, decorations.pathreplacing,
backgrounds, calc, positioning, fit, matrix}

\hypersetup{%
	pdfusetitle=true,%
	bookmarks=true,%
	pdftitle={The Case for Deep Query Optimisation},%
    pdfauthor={Jens Dittrich, Joris Nix},     % author
    pdfsubject={Databases},   % subject of the document
    pdfcreator={Jens Dittrich, Joris Nix},   % creator of the document
    pdfproducer={\LaTeX}, 
	pdfkeywords={Query optimisation, query planning, physical operators, systems},%
	pdfdisplaydoctitle=true,%
	pdfstartview=Fit,%
	colorlinks=false%
}

\definecolor{darkgreen}{rgb}{0.0, 0.38, 0}

\definecolor{b}{rgb}{0.0, 0.0, 1.0}
\definecolor{p}{rgb}{0.5019607843137255, 0.0, 0.5019607843137255}
\definecolor{r}{rgb}{1.0, 0.0, 0.0}
\definecolor{o}{rgb}{1.0, 0.6470588235294118, 0.0}
\definecolor{g}{rgb}{0.0, 0.5, 0.0}

\begin{document}
\sloppy

\title{The Case for Deep Query Optimisation\footnote{The authors apologise for using the adjective \textit{deep} in a paper title, in 2019!, even though this paper has nothing to do with \textit{deep} learning. We are deeply sorry.}}

\author{Jens Dittrich\qquad Joris Nix\\
\\
Big Data Analytics Group\\
Saarland University\\
Saarland Informatics Campus\\
\url{https://bigdata.uni-saarland.de}}

\date{\today}

\maketitle

\begin{abstract}
Query Optimisation (QO) is the most important optimisation problem in databases. The goal of QO is to compute the best physical plan under a given cost model. In that process, physical operators are used as building blocks for the planning and optimisation process. In this paper, we propose to deepen that process. We present Deep Query Optimisation (DQO). In DQO, we break up the abstraction of a `physical' operator to consider more fine-granular subcomponents. These subcomponents are then used to enumerate (sub-)plans both offline and at query time. This idea triggers several exciting research directions: (1)~How exactly can DQO help to compute better plans than (shallow) QO and at which costs? (2)~DQO can be used to precompute and synthesise database operators and any other database component as Materialised Algorithmic Views~(MAVs). (3)~We identify the Algorithmic View Selection Problem (AVSP), i.e.~which MAVs should be materialised when?

This paper presents the high-level idea of DQO using an analogy inspired from biology. Then we proceed to question the terms `physical' and `physical operator'. We present experiments with a `physical operator' formerly known as `hash-based grouping'. We benchmark that operator both independently as well as in the context of DQO-enabled dynamic programming. We conclude by sketching a DQO research agenda.
\end{abstract}

\section{Introduction}
Query Optimisation (QO)  is at the heart of any query engine. The core task of QO is to find an efficient plan under a given cost model. A major difficulty of this task is to compute not only a \textit{logical plan} (an extended relational algebra expression, a DAG of logical operators) specifying which relations to join in which order but additionally a \textit{physical plan} (a DAG of physical operators) that additionally specifies which access methods (e.g.~unclustered B-tree vs scan) and algorithms to use (e.g.~sort-merge vs hash join). The physical plan can then either be interpreted or compiled.

A major problem of this approach is the hidden legacy of relational algebra. For example, at some point in QO, a logical join, i.e.~$\bowtie\hspace*{-0.1cm}(R,S)$, is translated to a physical join, i.e.~$\texttt{SortMergeJoin}(R,S)$ or $\texttt{HashJoin}(R,S)$. In other words, the abstraction used during query optimisation is the following: a physical operator in relational algebra receives one or two inputs from outside, does some well-defined processing \textit{inside}, and produces one output dataset to the outside. In summary, a physical plan is DAG-structured `algorithmic recipe' where the nodes are physical operators and the edges symbolise producer-consumer relationships.

There has been considerable work on examining the pipelining aspects of this legacy, i.e.~how to effectively implement the producer-consumer relationships from good old volcano-style ONC-iterators~\cite{DBLP:journals/csur/Graefe93}, via vectorisation~\cite{DBLP:conf/cidr/BonczZN05} to breaking the boundaries of physical operators to run plans until the next pipeline breaker~\cite{DBLP:journals/pvldb/Neumann11}, to include parallelism~\cite{DBLP:conf/sigmod/LeisBK014}, and combinations thereof~\cite{DBLP:journals/pvldb/KerstenLKNPB18}.  

However, these works neglect that the algorithms used to implement an operator \textit{can be considered a query plan in itself}. For instance, consider the physical grouping operator\footnote{We focus on grouping here as joins have very similar algorithmic issues and solutions. This becomes clear if you consider that a join is merely a co-group-operation with exactly two inputs followed by an aggregation on each co-group.Vice versa a grouping operation is merely a co-group operation with a single input.}. From a 10,000 feet perspective, it can be implemented either using a sort-based- or a hash-based algorithm. 
In hash-based grouping (as shown in Figure~\ref{fig:hbgrouping}) we initialise an empty hash table (Line~1). Then we insert each tuple from the input into that hash table using the grouping key as the key to the hash table and the set of tuples having that key as the value to the hash table (Lines~2--6). Afterwards, for each existing key in the hash table, we compute the aggregate on the set of tuples pointed to (Lines~7--8).

\begin{figure}[h]
\begin{center}
\begin{footnotesize}
\fbox{\parbox{.5\textwidth}{
\textbf{HashBasedGouping (Relation R, groupingKey):}
\begin{enumerate}[itemsep=0pt]
\item HashMap hm; Relation result = \{\};
\item [] \hspace*{-0.45cm}//Insert all tuple from input R into HashMap hm:
\item  For each  r in R:
\item  \tabto{.7cm} If  r.groupingKey in hm:
\item  \tabto{1.5cm} hm.probe(r.groupingKey) $\cup=$ \{r\};
\item  \tabto{.7cm} Else:
\item  \tabto{1.5cm} hm.insert(r.groupingKey, \{r\});
\item [] \hspace*{-0.45cm}//Build aggregates for each existing key in hm:
\item  For each  key in hm.keySet():
\item  \tabto{.7cm} result $\cup$= aggregate(hm.probe(key));
\item  Return result;
\end{enumerate}}}
\end{footnotesize}
\vspace*{-0.2cm}
\caption{\label{fig:hbgrouping}Textbook-style pseudo-code for hash-based grouping}
\end{center}
\end{figure}

This algorithm can be found in almost all database textbooks, lectures, and even conference talks. The problem with this algorithm is that it \textit{implies a couple of algorithmic and physical design decisions}:
\noindent(1)~As an internal index structure a hash table is used, but which one exactly? As already observed
    in~\cite{DBLP:journals/pvldb/0007AD15} a hash table has many different dimensions which influence performance dramatically.
\noindent(2)~The insert operations to the hash table are implicitly assumed to occur in serial.
\noindent(3)~The aggregation operations in the hash table are implicitly assumed to occur in serial, i.e.~group-wise.
\noindent(4)~The `Relation R' parameter in the function signature implies that the entire result set is passed to the algorithm fully materialised. Likewise, the final aggregates are collected in a result set before passing them outside which again implies that the result set is materialised. In summary, this induces two unnecessary pipeline breakers.
\noindent(5)~The two phases in the algorithm (first load the hash table, then compute the aggregates) forbids any kind of non-blocking behaviour, e.g.~like in any kind of online aggregation algorithm~\cite{DBLP:conf/sigmod/HellersteinHW97, DBLP:conf/vldb/DittrichSTW02}.

A much better description of this algorithm is shown in Figure~\ref{fig:PartitionBasedGrouping}.
The two LOCs in Figure~\ref{fig:PartitionBasedGrouping} basically say: (1)~we will partition the data produced by R into a bundle of independent producers. If the input produces 42 different groups, partitionBy creates 42 different producers. Semantically each producer will deliver the tuples belonging to its group. Notice that there is no need here to shoehorn the result into one relation as in relational algebra or SQL. In (2)~we specify that a bundle of independent producers is aggregated with the same aggregation function, but possibly independently.

\begin{figure}[h!]
\begin{center}
\includegraphics[trim = 10mm 65mm 65mm 12mm, clip, width=\textwidth,height=\textheight,keepaspectratio,page=1]{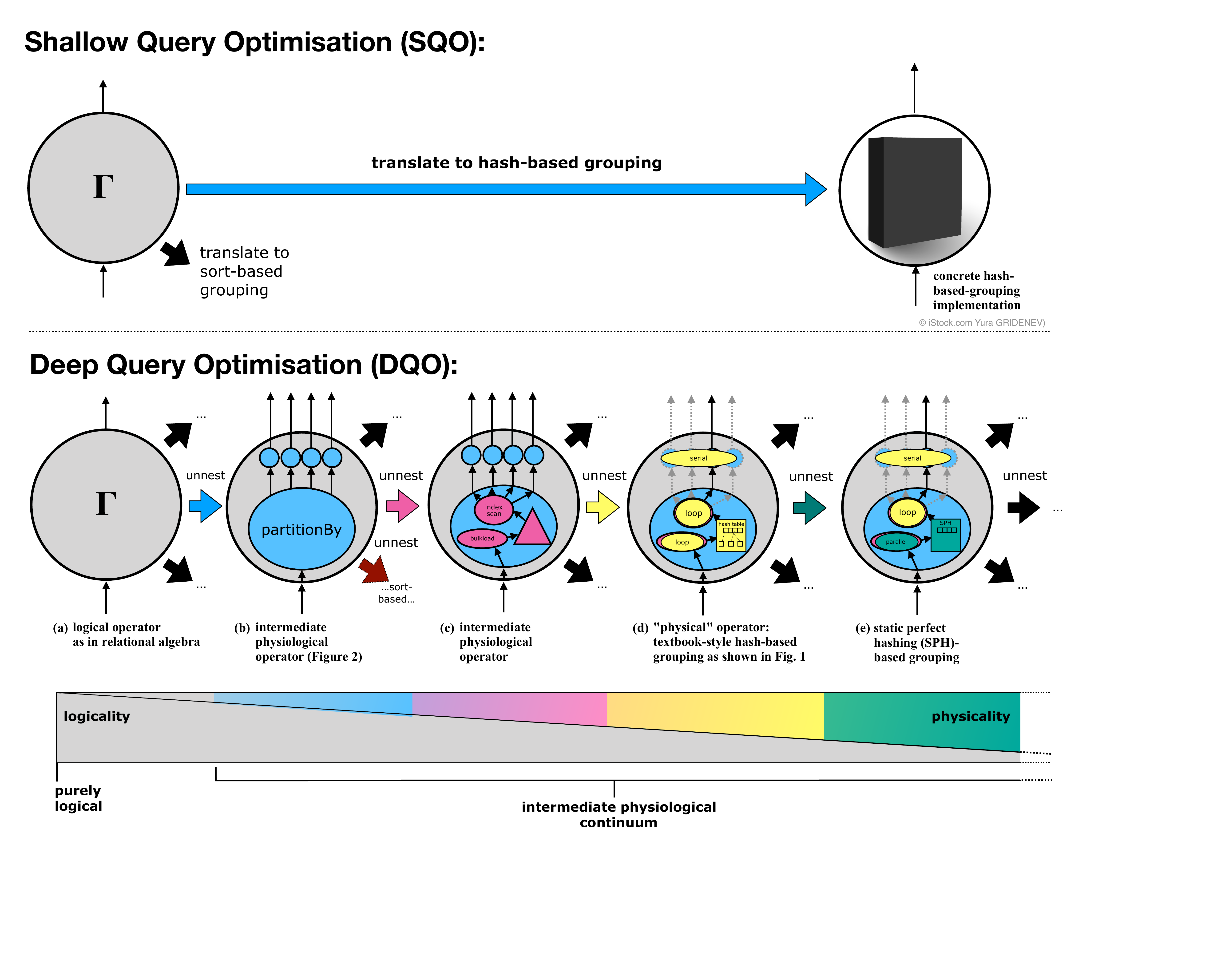}
\end{center}
\vspace*{-0.2cm}
\caption{\label{fig:unboxing}
Standard (`shallow') query optimisation (SQO) vs deep query optimisation (DQO)
Top: Different unnesting steps of the logical grouping operator: each colour depicts yet another hidden nested query plan. Observe that every time we unnest, we make certain algorithmic decisions. Thus, \textit{each unnest operation increases the physicality of the plan}. For instance, if we decide to implement partitionBy using any type of index, that decision will exclude other options like sort-based grouping. In turn, if we decide to implement that index using a hash table, we exclude other types of indexes. Notice that naming a plan `physical' is actually misleading, as there is almost always yet another translation step underneath. For instance, the plan in (e) increases the physicality by choosing one particular type of hash table, hash function, etc (SPH in this case). In further unnesting steps (not shown), code compilation by the database system and/or compiler will add more steps. Finally, compiler and hardware will make more decisions, e.g.~reordering of code and memory accesses. Probably, only the final actions performed on the hardware (the \textit{physics}) should be called a truly \textit{physical plan}.
\newline Bottom: the design space continuum starting on the left from purely logical to plans increasing the physicality of the plan. So far, we have mostly ignored the intermediate physiological plans in that middle-ground: that is what we consider in Deep Query Optimisation~(DQO).
}
\end{figure}

What if we depict this `code', i.e.~the insides of the logical operator, as a query plan itself? This is shown in Figure~\ref{fig:unboxing}. Figure~\ref{fig:unboxing}(a) shows logical grouping as found in (extended) relational algebra. If we open up that box, i.e.~if we \textit{unnest} the operator into a more fine-granular query plan, we obtain Figure~\ref{fig:unboxing}(b). The latter corresponds to the pseudo-code given in Figure~\ref{fig:PartitionBasedGrouping}. The more we unnest, we \textit{increase the physicality} of the query plan as for each unnest we have to make a decision on \textit{how exactly} to implement a certain `bubble'. Only after some recursive unnesting, we eventually obtain Figure~\ref{fig:unboxing}(d) which corresponds to the pseudocode of hash-based grouping shown in Figure~\ref{fig:hbgrouping}. In other words, hash-based grouping is just one of many special cases in a partition-based grouping algorithm. Figure~\ref{fig:unboxing}(e) shows another unnest using static perfect hashing (SPH, in the experiments we will even use minimal SPH) as well as a parallel load. And from that we could continue further...

In Figure~\ref{fig:unboxing}, the  arrows denote a specific path we followed at each unboxing step. In turn, at each unnest we discard several options. In this example we unnest four times. However, this figure just visualises the principle. We do not imply that it is exactly four unnest operations to get from a logical operator to a `physical' plan. However, what \textit{we do imply is that the current state-of-the-art to translate from one extreme (a logical operator) in a single step to another extreme (a blackbox `physical operator') misses several interesting optimisation opportunities} (cf.~SQO in Figure~\ref{fig:unboxing}).

\begin{figure}[t]
\begin{center}
\begin{footnotesize}
\fbox{\parbox{.7\textwidth}{
\textbf{PartitionBasedGrouping(Producer R, Consumer R', groupingKey):}
\begin{enumerate}[itemsep=0pt]
\item [] \hspace*{-0.45cm}//partition the input:
\item R $\rightarrow$ partitionBy(groupingKey) $\rightrightarrows$ R\_partitions
\item [] \hspace*{-0.45cm}//aggregate each partition:
\item R\_partitions $\rightrightarrows$ aggregate(...) $\rightrightarrows$ R'
\end{enumerate}}}
\end{footnotesize}
\vspace*{-0.2cm}
\caption{\label{fig:PartitionBasedGrouping}Partition-based grouping. Here each `line of code' is written as a producer-consumer pattern, i.e.~a line of code consumes some input and creates one or multiple producers. This does not make \textit{any} algorithmic decision whatsoever on how this producer-consumer pattern will be implemented physically. $\rightrightarrows$ denotes that an operation provides a bundle of independent producers.}
\end{center}
\end{figure}

This paper is structured as follows: in the next section we introduce Deep Query Optimisation (DQO). In Section~\ref{sec:algorithmicviewselection} we introduce Materialised Algorithmic Views (MAVs) and the Algorithmic View Selection Problem (AVSP). Section~\ref{sec:systemintegration} discusses how to integrate DQO into existing systems.
Section~\ref{sec:experiments} presents early experimental results of our idea. Section~\ref{sec:relatedwork} contrasts DQO to related work. Finally, Section~\ref{sec:researchagenda} presents a research agenda.

\section{Deep Query Optimisation}
The core idea of \textbf{Deep Query Optimisation (DQO)}
is the following: rather than computing a plan using only coarse-granular logical and/or `physical' operators (as done in Shallow Query Optimisation, SQO), in DQO we consider more fine-granular components in the optimisation process.

As already shown in the previous sections, when inspecting a single group-by operator, there are many different hidden levels of nested query plans. Table~\ref{tab:granularities} presents another view on these granularity levels using an example from biology, a living cell, as an analogy. A living cell is composed of organelles which consist of (macro-)molecules which consist of atoms. Using the living cell analogy, we can phrase the key idea of DQO as follows: \textit{extend SQO to \textbf{also} assemble organelles \textit{and} macro-molecules from molecules rather than only living cells from organelles.}

Table~\ref{tab:SQOvslearnedvsDQO} contrasts SQO, and machine-``learned'' techniques~\cite{DBLP:journals/sigmod/0059Z0JOT16} with DQO.

\begin{table}[h!]
\begin{center}
    \includegraphics[width=\textwidth]{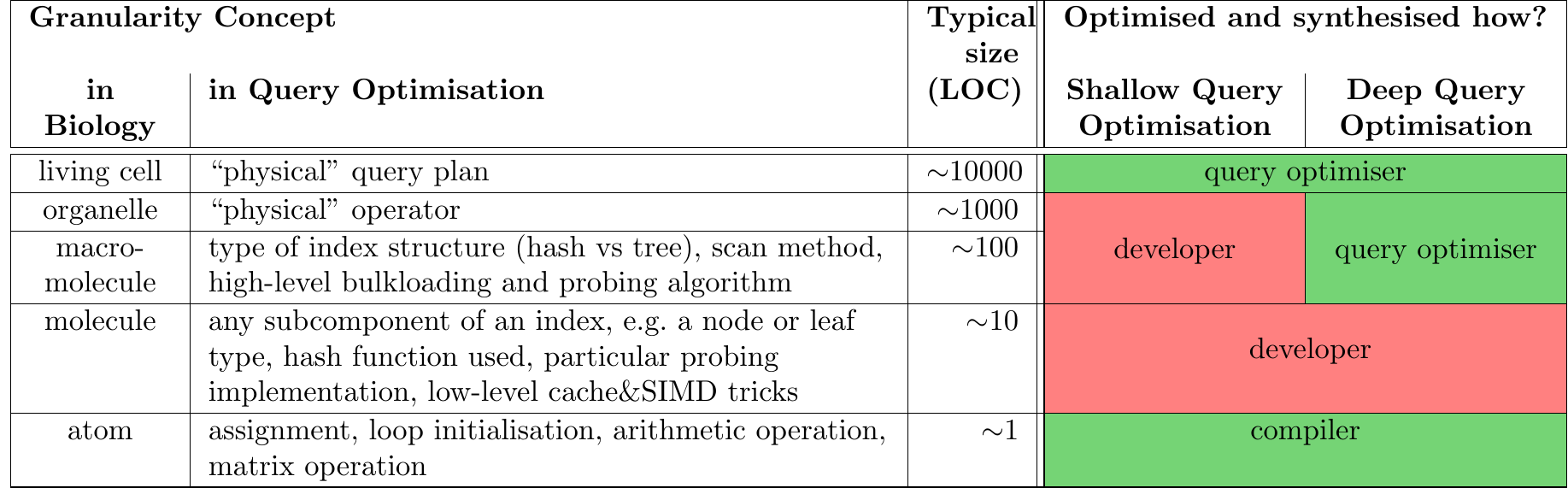}
\end{center}
\vspace*{-0.3cm}
\caption{\label{tab:granularities}
Granularity concepts in biology vs their counterparts in query optimisation, their typical sizes, and example instances. The main difference of DQO over SQO is that we push the frontier of what can be optimised further down into `physical' operators. Hence, a physical operator is not anymore a given thing optimised manually by some developer, but optimised and synthesised by the system, either at query time or beforehand.
}
\end{table}

\begin{table}[h!]
\begin{center}
    \includegraphics[width=1.0\textwidth]{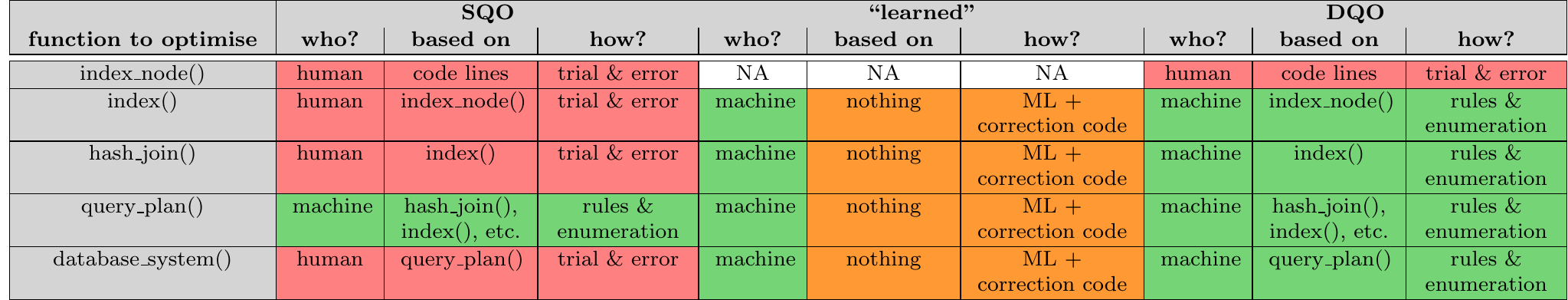}
    \end{center}
\vspace*{-0.3cm}
\caption{\label{tab:SQOvslearnedvsDQO}
SQO, ``learned'', and DQO in comparison: both SQO and DQO assume that we rewrite and enumerate possible plans. However in contrast to SQO, in DQO we assume components at a much finer granule (as shown in Table~\ref{tab:granularities}).  In contrast to ``learned'' approaches, in DQO, we do not learn those functions from scratch but assume fine-granular building blocks to be available. Similar to SQO, we assume that rules are available that define how to combine those building blocks, i.e.~we see this primarily as an optimisation rather than a machine learning problem.  Another important difference and advantage over several ``learned'' approaches and a similarity to SQO is that we do not need extra mechanisms to correct errors due to the approximative nature of most ML-methods.
}
\end{table}

\subsection{Local vs Global Effects of Deep Query Optimisation}
DQO can be considered \textit{local} if the types of subcomponents used to assemble a specific granule do not have any effect on the context of that granule. 
In contrast, DQO must be considered \textit{global} (or at least non-local) if the types of subcomponents used to assemble a specific granule do (or even may) have an effect on the context of that granule. 

For instance, assume we want to find the optimal index to be used in hash-based grouping. Let's take a look at two major options for the indexes to use here:

\noindent(1)~We use an out-of-the-box hash-table. Then, the output of the operator will not be sorted (technically it is often sorted in the order as the groups appear in the hash table. That order depends heavily on the hash function used. If we do not know exactly which order is produced by a blackbox hash table, we have to assume that the data is unordered to be on the safe side.). 

\noindent(2)~We use a static perfect hash-function (SPH). SPH can simply be an array of groups of tuples (or running aggregates in the case of distributive and/or decomposable aggregation functions). The grouping key then serves as the index into that array. Here, the linear array slot computation works like a perfect hash function. If all array slots are used, the SPH is even minimal. This is only applicable if the key domain of the grouping key is (relatively) dense. This situation is not as rare as one might think. For instance, the keys of a dictionary-compressed column are a natural candidate for this and can directly be used for SPH. Like in a hash-table, the grouped output will be sorted according to the order in the underlying array.
Notice that in both options,  the order on the probe input is preserved.

In summary: any optimisation step in DQO can be considered local, if the subplan produced at that step has the same properties as any other plan at that granule.
In contrast if, for a given granule different subplans have different properties, its optimisation effects may be considered global. 

\subsection{Meta-Relational Plan Properties}
\label{sec:DQOplanproperties}
Throughout the years, database research literature proposed concepts like \emph{interesting orders} in sort-based operators or physical properties like \emph{compressed} or \emph{partitioned} data ~\cite{DBLP:conf/icde/GraefeM93}.  
However, in DQO we do not limit ourselves to this narrow set of plan properties and expand the \emph{property space}.  Therefore, we introduce the concept of \emph{meta-relational properties}, i.e.~properties that go beyond a pure relational model and reasoning about a query plan.  Table~\ref{tab:meta-rel-prop} gives an overview of existing meta-relational properties including a first attempt to categorise them into various sub-categories. Concepts like \textit{interesting orders} are included as a special-case.

\begin{table}[h]
    \centering
    \begin{tabular}{>{\centering\arraybackslash}p{2.7cm}|>{\centering\arraybackslash}p{2.3cm}|>{\centering\arraybackslash}p{2.3cm}}
        \textbf{physical}    & \textbf{structural}    & \textbf{statistical}   \\\hline\hline
        data layout      & partitioning  & distribution  \\
        compression & grouping      & cardinality   \\
        location    & sorting       & uniqueness    \\
        hardware    & correlation   & density     \\
        clustered   &               & \# rows     \\
                    &               & \# columns     \\
                    &               & \# NULL-values     \\
                    &               & min, max, avg, etc.     \\
    \end{tabular}
    \vspace*{-0.2cm}
    \caption{\label{tab:meta-rel-prop}Meta-relational properties for deep query optimisation.}
\end{table}
\noindent\textbf{Physical} meta-relational properties are those properties that are abstracted away by physical data independence, e.g.~any kind of data layout, physical location, and/or internal representation of the data.

\noindent\textbf{Structural} meta-relational properties are related to ordering and/or partitioning of the data, e.g.~\textit{sorting} is referred to in database literature as \textit{interesting order}. As these properties do not necessarily have to be represented physically they belong to a separate category. Still they \textit{may additionally} also be reflected physically in which case they belong to the category \textit{Physical meta-relational properties}.

\noindent\textbf{Statistical} meta-relational properties are statistical properties summarising the data using statistical moments or any other useful (yet lossy) approximation.

As optimisation in DQO happens on small granules, we have to identify the right set of meta-relational properties that allow for a deeper optimisation on the corresponding granule, e.g.~on an operator level or on the structure of the query plan. In addition, we have to consider that many of these meta-relational-plan properties may have a non-local effect. 

In a first shot, the properties with a non-global effect can be considered and handled very similarly to how interesting orders are handled in dynamic programming. If any subcomponent in DQO produces an output with such a property, we \textit{must not} discard that information.

In the experiments in Section~\ref{sec:experiments}, we show how we modified dynamic programming to factor in the following meta-relational plan property: dense vs sparse and how that makes a difference in optimisation and runtime performance.  Examining the other properties in more detail is an avenue for future work.

\section{Materialised Algorithmic View Selection}
\label{sec:algorithmicviewselection}

\begin{figure}[h!]
\begin{center}
    \includegraphics[trim = 0mm 0mm 0mm 0mm, clip, width=.97\textwidth,height=\textheight,keepaspectratio,page=1]{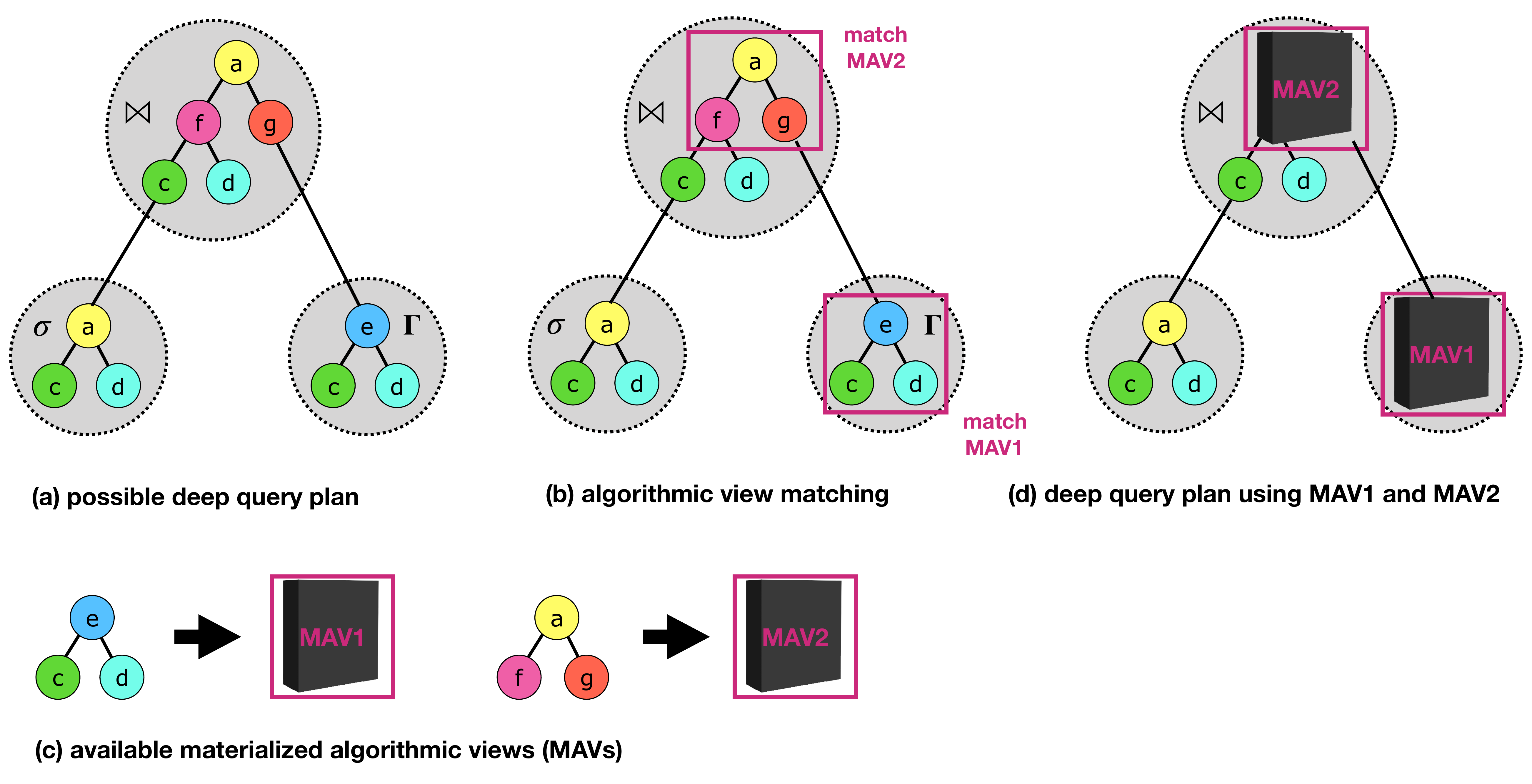}
\end{center}
\vspace*{-0.3cm}
\caption{\label{fig:avsp}Materialised Algorithmic Views (MAVs) in deep query planning and  the Algorithmic View Selection and matching Problem (AVSP)
}
\end{figure}

There is another interesting similarity of DQO to SQO when recalling materialised views (MVs, precomputed query results) and also prepared statements (preoptimised queries).

\subsection{Materialised Algorithmic Views (MAVs)}

In DQO, in particular for local effects but also for the non-local ones, it makes sense \textit{to precompute certain granules offline (before a query comes in)}. We coin these precomputed components \textbf{Materialised Algorithmic Views (MAV)}. Just like MVs prematerialise query results, MAVs prematerialise (i.e.~optimise) parts of a deep query plan. An MAV may be a component of a physical ``operator'', entire physical ``operators`` or larger units, i.e.~physical subplans.

In other words, a deep query plan as shown for instance in Figures~\ref{fig:unboxing}(b--e) may conceptually represent (parts of) an algorithm, an index node/leaf, an index, a physical operator like a very specific hash join algorithm implementation or even larger parts of a physical plan. However, as the search space for this plan is potentially large, we believe that MAVs may help in that they materialise certain subtrees of that plan into preoptimised units (again: just as prematerialised views are used as building blocks in shallow query optimisation).

MAVs can be precomputed for any granularity level not only within the traditional boundaries of an operator. Like that MAVs can be used as building blocks for DQO at query time to speed-up plan enumeration.

\subsection{When to materialise MAVs?}
An interesting question is when to compute an MAV. Again, the duality to materialised view selection helps a lot here: MVs work well if the underlying data does not change too often and we do not spend too much work updating those MVs. In addition, an MV should be useful in queries in the sense that it is used (matched) regularly and decreases the overall query runtime (including query optimisation time). At the same time the MV should be storage efficient. For MVs these trade-offs naturally lead to a cost/benefit-calculation to decide which MVs to keep. We believe that very similar techniques will strike for MAVs. We envision a  usefulness-ranking which quickly identifies components of a deep query plan that can be reused over and over again. This is up to future work to show.

There is a natural trade-off here: how much time do I want to spend on DQO offline vs at query time?

\subsection{The Algorithmic View Selection Problem (AVSP)}
At query time we have to face the challenge of matching those precomputed MAVs to suitable subplans.  Again, inspired by the materialised view selection problem~\cite{DBLP:conf/vldb/BaralisPT97}, we coin this the \textbf{Algorithmic View Selection Problem (AVSP)}. Like with MVs, there is no need in AVSP to make any manual decision about which granules to precompute. This decision can be automated. This is simply adding a new AVSP-dimension to the physical design problem.  

An example for algorithmic view matching is shown in Figure~\ref{fig:avsp}.  Figure~\ref{fig:avsp}(a) shows an example of a fine-granular (deep) query plan that we need to optimise. In Figure~\ref{fig:avsp}(b) we match the available MAVs as listed in Figure~\ref{fig:avsp}(c) against the plan in Figure~\ref{fig:avsp}(a). In Figure~\ref{fig:avsp}(b), we visualise two matches: both MAV1 and MAV2 can be matched against suitable subplans. In Figure~\ref{fig:avsp}(d), we replace these matches with their corresponding preoptimised MAVs. Only after this matching step, plan enumeration needs to be invoked. Hence, the search space is diminished --- just like in materialised view selection. Figure~\ref{fig:avsp} also shows the relationship to MVs: an MV for a given subplan precomputes a query result. This only works if that subplan corresponds to a subtree of the plan, i.e.~it does not depend on additional computations that need to be performed at query time. In contrast, an MAV materialises an algorithm. Still that algorithm may rely on inputs of other algorithms which are only optimised at runtime. In that sense, an MAV is also related to prepared statements which partially preoptimise a plan but still allows the database engine to insert variable assignments. In that sense, in order to map these ideas back to the traditional database universe of logical and physical operators, an MAV that happens (incidentally) to prematerialise the internals of what we used to call an operator could be coined a \textbf{prepared operator}. For instance, in Figure~\ref{fig:avsp}(d), MAV1 materialises the contents of a grouping operator. Again, MAVs do not have to match these traditional operator  boundaries. In Figure~\ref{fig:avsp}(d), the join operator is an example for a case where only parts of the ``operator'' are replaced by MAV2.

Another relationship worth mentioning is plan caching: some database engines already keep subplans of previously optimised queries and reuse them. However, in contrast to DQO, these techniques work on the coarse-granular operator abstraction. 

Also notice that there is \textbf{no need to fully preoptimise} an MAV. A ``materialisation'' simply means that we decrease the logicality and increase the physicality of this subplan in whatever way. Recall our discussion for Figure~\ref{fig:unboxing} which displays this effect as well: prematerialising an MAV simply means that we push a (sub-)plan further to the right.

MAVs and the AVSP trigger a couple of interesting research challenges and directions, in particular when applied to indexing. We will discuss them in Section~\ref{sec:researchagenda}.

\subsection{But the Search Space is Exponential!}
It is obvious that DQO increases the search space of an already (but only possibly!) exponential search space even more.
However, in SQO, it is well-known that the shape of the join graph may reduce the complexity of the search space dramatically, e.g.~the search space is only polynomial for linear plans,  see~\cite{Moerkotte2006BuildingQC} for an overview.  We believe that similar effects are present in DQO, i.e.~the shape of the graph representing how components may be composed may reduce the search space a lot. This is up to future work to explore.

We also believe that similarly to how MVs helped to reduce the search space of a shallow query plan at query time MAVs help to reduce the search space over algorithmic components. 
So, in summary, the explosion of the search space may seem frightening in the beginning, however, that is not a reason to  ignore DQO but rather a call to research.

\section{System Integration}
\label{sec:systemintegration}
Our experiments in Section~\ref{sec:experiments} consider the effects of DQO in an isolated way by examining a concrete example of different grouping implementations. The intention is to make the case for a deeper level of query optimisation without having a complete system behind it. Our longterm vision is to integrate this core idea into an existing database management system. Therefore, in this section we will discuss what the integration of DQO into an existing DBMS could look like. For this, we will briefly touch on the core components of query optimisation like plan representation, plan space enumeration, cost models, and statistics. Ultimately, we are planning to integrate DQO into mu\emph{t}able~\cite{mutable}, a query execution engine currently developed at Saarland University.\\ \noindent\textbf{Plan representation.} The most fundamental design decision, on which the other parts of the query optimisation engine are building up, is the internal representation of a query. Since we aim to break up the current structure of operators and to optimise on a deeper level, a new physiological representation of queries might be necessary (analogue to one of the intermediate physiological operator representations in Figure~\ref{fig:unboxing}). For instance, instead of having a logical grouping operator, we could introduce multiple physiological grouping operators which already decide on some implementation details like index-based or order-based grouping (see our algorithms in Section~\ref{sec:experiments}).

\noindent\textbf{Plan space enumeration.} The deeper we go into an operator, the more the search space and the complexity increases. Holistically enumerating the whole query plan including all possible subcomponents of every operator could easily exceed what is computationally possible right now. To overcome this, one possibility might be to do independent plan enumeration inside each operator using the structural components as the building blocks. With that, we can locally come up with the best possible operator given the underlying dataset and workload. Once we have the optimal physical operators, we can use classical plan enumeration algorithms like dynamic programming to find the final query execution plan. Of course, this probably does not find the overall best execution plan because we are missing global optimisation effects. However, this could already improve performance significantly. To further prune the search space, we can fall back on materialised algorithmic views. We could essentially create a library of many different algorithmic implementations from which we could choose.  

\noindent\textbf{Cost models.} Which cost models we are effectively going to use depends on the type of plan enumeration and which statistics are available to use. In case we want to enumerate the structure of operators, we might need separate cost models for the different kinds of operators. In addition, we expect to come across new rules which we can incorporate into the optimisation process. For example, using distribution information by knowing that the underlying data is sparse, we might discard all possible plans using static perfect hash-based algorithms because the performance will definitely be worse than classical hash-based implementations (see our experimental results in Section~\ref{sec:experiments}).  

\noindent\textbf{Statistics.} To enable a fine-granular optimisation of our query operators and with it, of the complete query plan, we need precise statistics about the underlying data and the workload. For queries processing large amounts of data, we assume the overhead for collecting additional statistics to be most likely negligible. Which statistics we need depends on the optimisation decisions we want to make. For example, information about the density of the data allows the use of specialised algorithms like static perfect hash-based grouping.

As a closing thought, we expect that many design decisions will come naturally as we are implementing the system.

\section{Experiments}
\label{sec:experiments}
DQO can be applied in all parts of query optimisation that include algorithmic design decisions.  In this section, we demonstrate in a small domain that DQO can have a significant impact on query execution.

\subsection{Setup and Methodology}
All experiments were conducted on a Linux machine with an AMD Ryzen Threadripper 1900X 8-Core processor with 32~GiB memory. All algorithms are implemented single-threaded in C++ and compiled with Clang 8.0.1, -O3. 

We consider five different implementation variants of grouping\footnote{The implementation of our algorithms is available on GitHub~\url{https://github.com/BigDataAnalyticsGroup/Deep-Query-Optimization} }. Each implementation computes the aggregates COUNT and SUM on the fly and stores a mapping from grouping key to aggregate data inside an array. In our experiments, we assume the number of distinct values to be known. However, this assumption is not strictly required and can be relaxed by using dynamic data structures, techniques like rehashing, or arrays that cover the whole domain in case it is relatively small and dense.

\noindent\textbf{Hash-based Grouping (HG).} We use \texttt{std::unordered\_map} as the underlying hash table and the Murmur3 finaliser as hash function. Every input element is inserted individually into the hash table.

\noindent\textbf{Static Perfect Hash-based Grouping (SPHG).} We use the grouping key as offset into the array storing the groups, acting as a static and perfect hash function.

\noindent\textbf{Order-based Grouping (OG).} This implementation requires the input data to be partitioned by the grouping key.  We iterate sequentially over the input data, create a group for the very first occurrence of a grouping key, and insert this group at the first empty slot in the array. As long as the grouping key remains the same, the corresponding aggregates are updated.

\noindent\textbf{Sort \& Order-based Grouping (SOG).} We do not require that the input data is partitioned by the grouping key. Therefore, we first sort the data then we apply OG.

\noindent\textbf{Binary Search-based Grouping (BSG).} We  store a mapping from grouping key to aggregate data inside a sorted array. This allows us to perform binary search to lookup a group by its key.

The datasets consist of 100 million 4 byte unsigned integer values representing the grouping key. Each dataset is uniformly distributed and has two properties, sortedness and density. Taking all combination of those properties, we end up with four different datasets.

\subsection{Performance of Physical Grouping Implementations}
\label{subsec:perf-grp}
\begin{figure}[h!]
\begin{center}
    \begin{tikzpicture}[node distance=-\pgflinewidth, rec/.style={rectangle, inner sep=0pt, minimum width=3mm, minimum height=1mm},
        font=\scriptsize]
        \node[rec, fill=b] (HG) at (0,0) {};
        \node (HGLabel) [right=of HG] {Hash-based Grouping};
        \node[rec, fill=r, node distance=2mm] (OG) [right=of HGLabel] {};
        \node (OGLabel) [right=of OG] {Order-based Grouping};
        \node[rec, fill=o, node distance=2mm] (SOG) [right=of OGLabel] {};
        \node (SOGLabel) [right=of SOG] {Sort \& Order-based Grouping};
        \node[rec, fill=g, node distance=2mm] (SPHG) [below=of HG] {};
        \node (SPHGLabel) [right=of SPHG] {Static Perfect Hash-based Grouping};
        \node[rec, fill=p, node distance=2mm] (BSG) [right=of SPHGLabel] {};
        \node (BSGLabel) [right=of BSG] {Binary Search-based Grouping};
    \end{tikzpicture}

\begin{tabular}{p{0.5cm}|c|c}
    & \textbf{sparse} & \textbf{dense}\\\hline
      \rotatebox{90}{\textbf{sorted}} &
    \raisebox{-.35\height}{\includegraphics[trim=0 0 0 -2,width=0.4\textwidth]{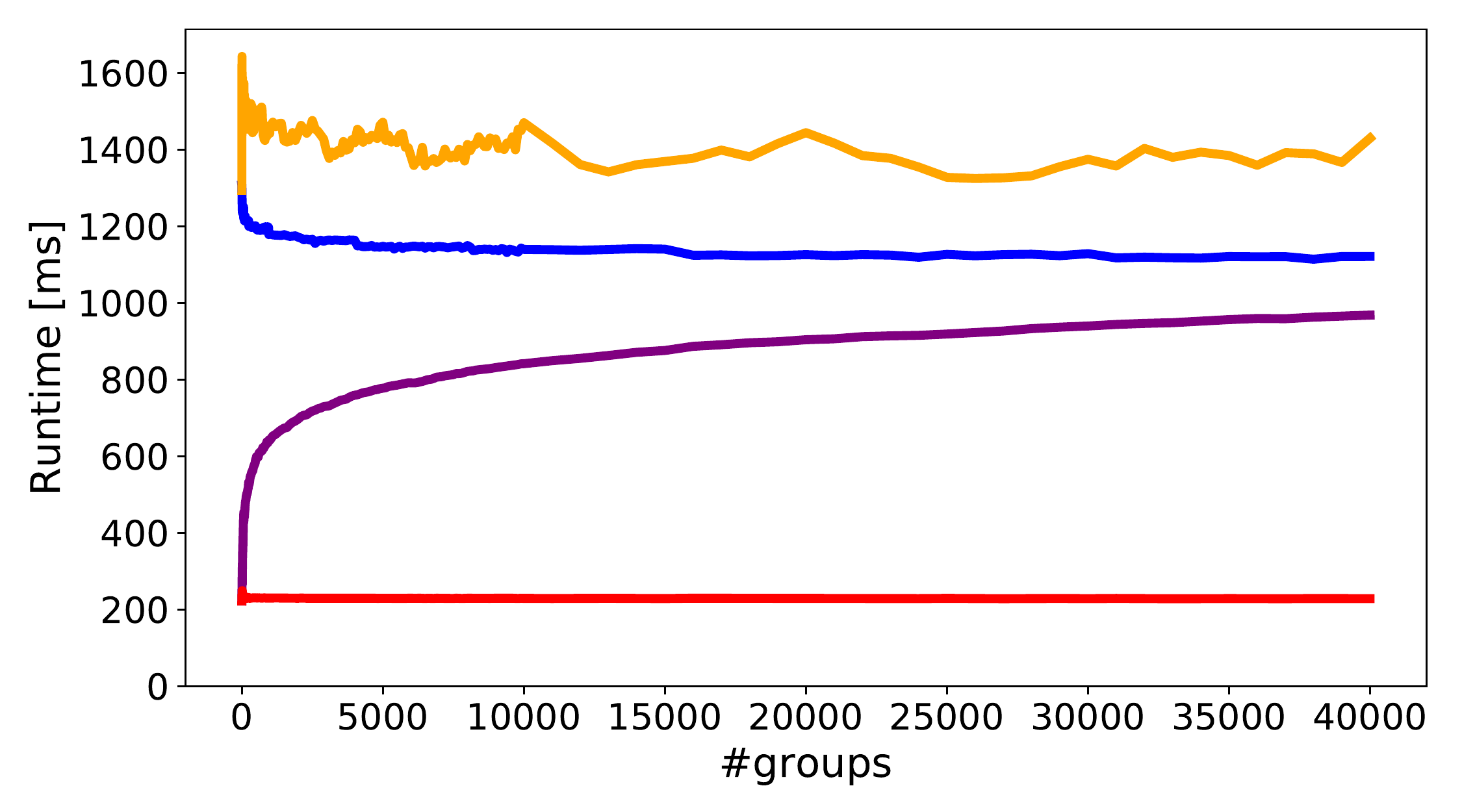}} &
    \raisebox{-.35\height}{\includegraphics[trim=0 0 0 -2,width=0.4\textwidth]{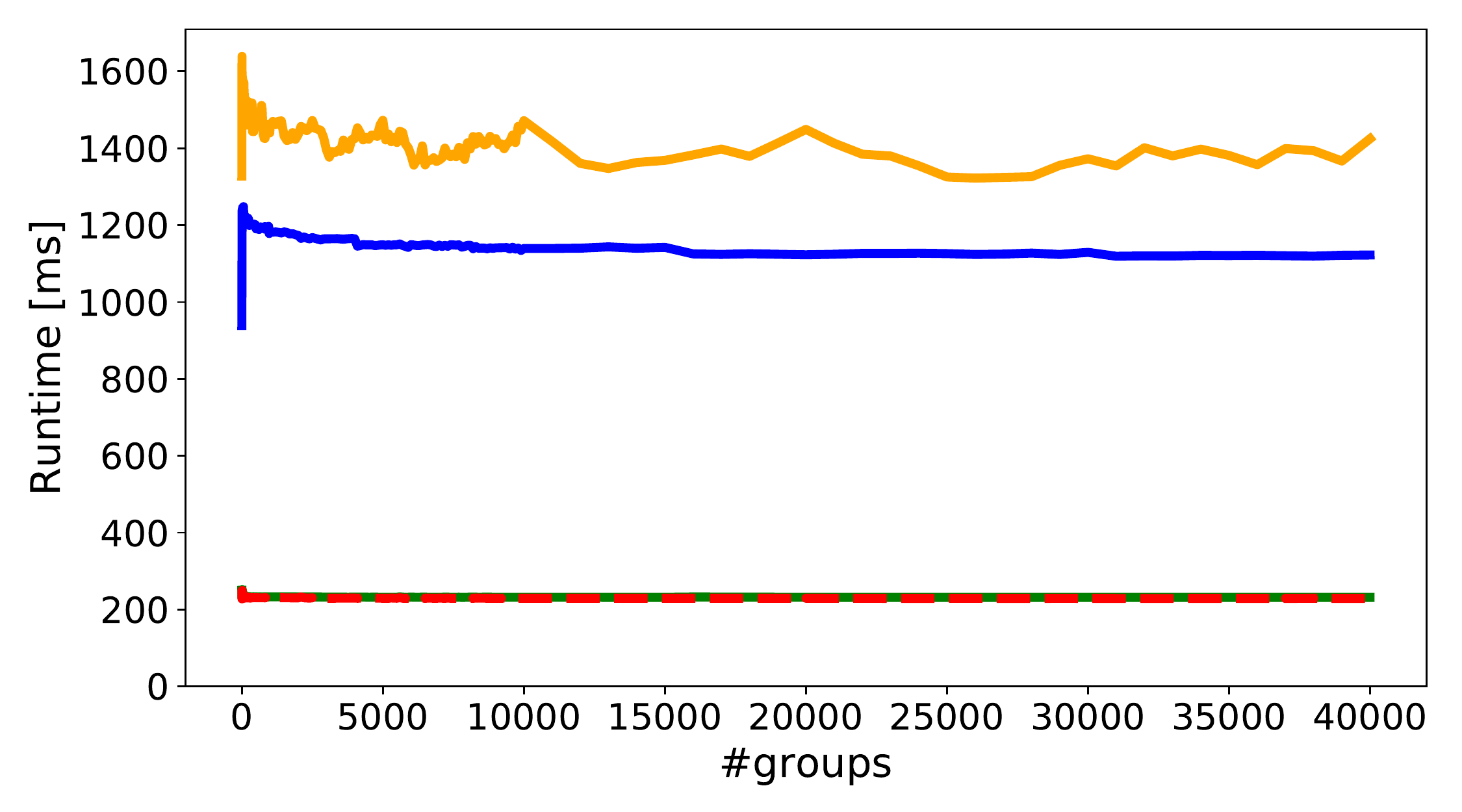}}   \\\cline{1-3}
     \rotatebox{90}{\textbf{unsorted}}  &
    \raisebox{-.35\height}{\includegraphics[trim=0 0 0 -2,width=0.4\textwidth]{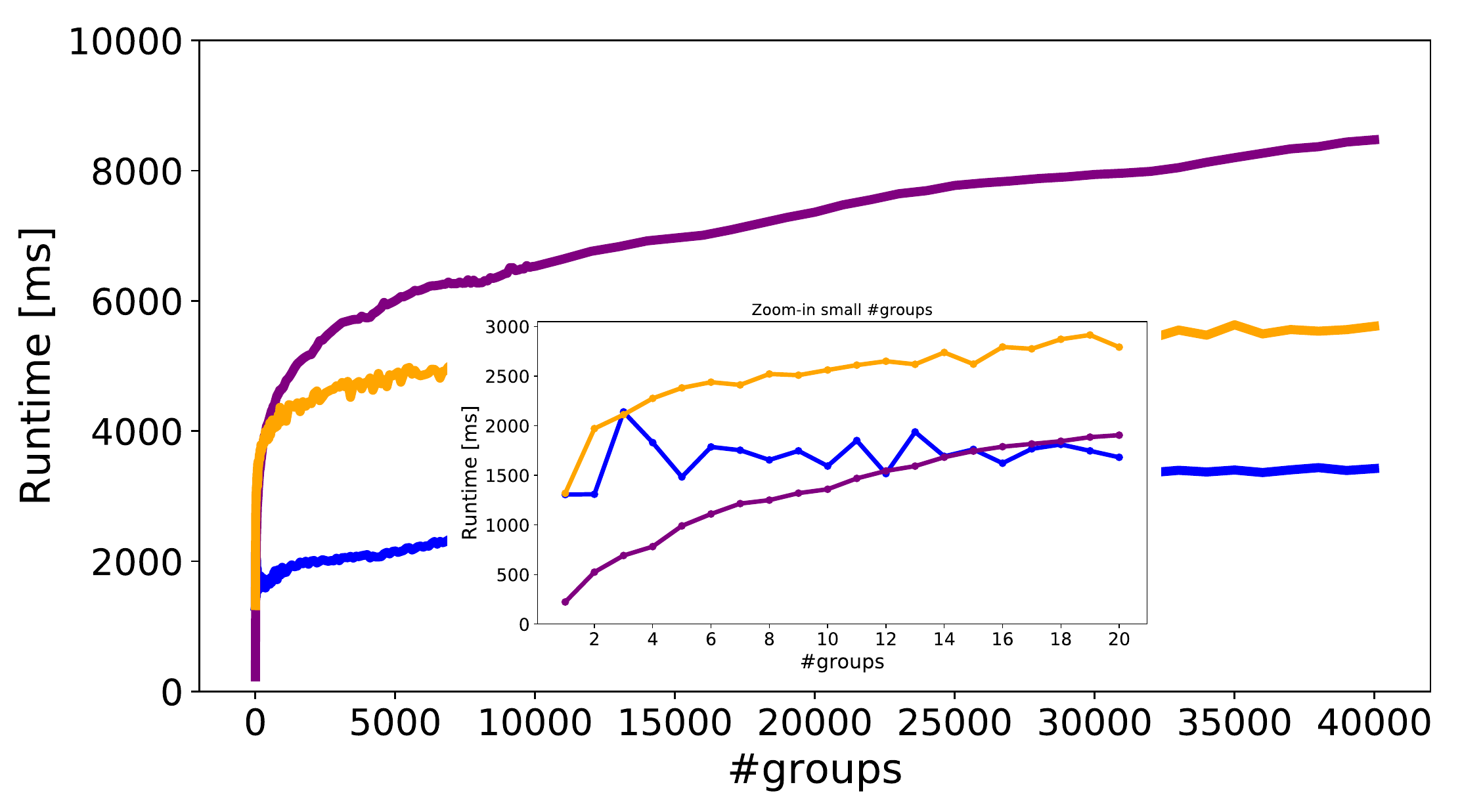}} &
    \raisebox{-.35\height}{\includegraphics[trim=0 0 0 -2,width=0.4\textwidth]{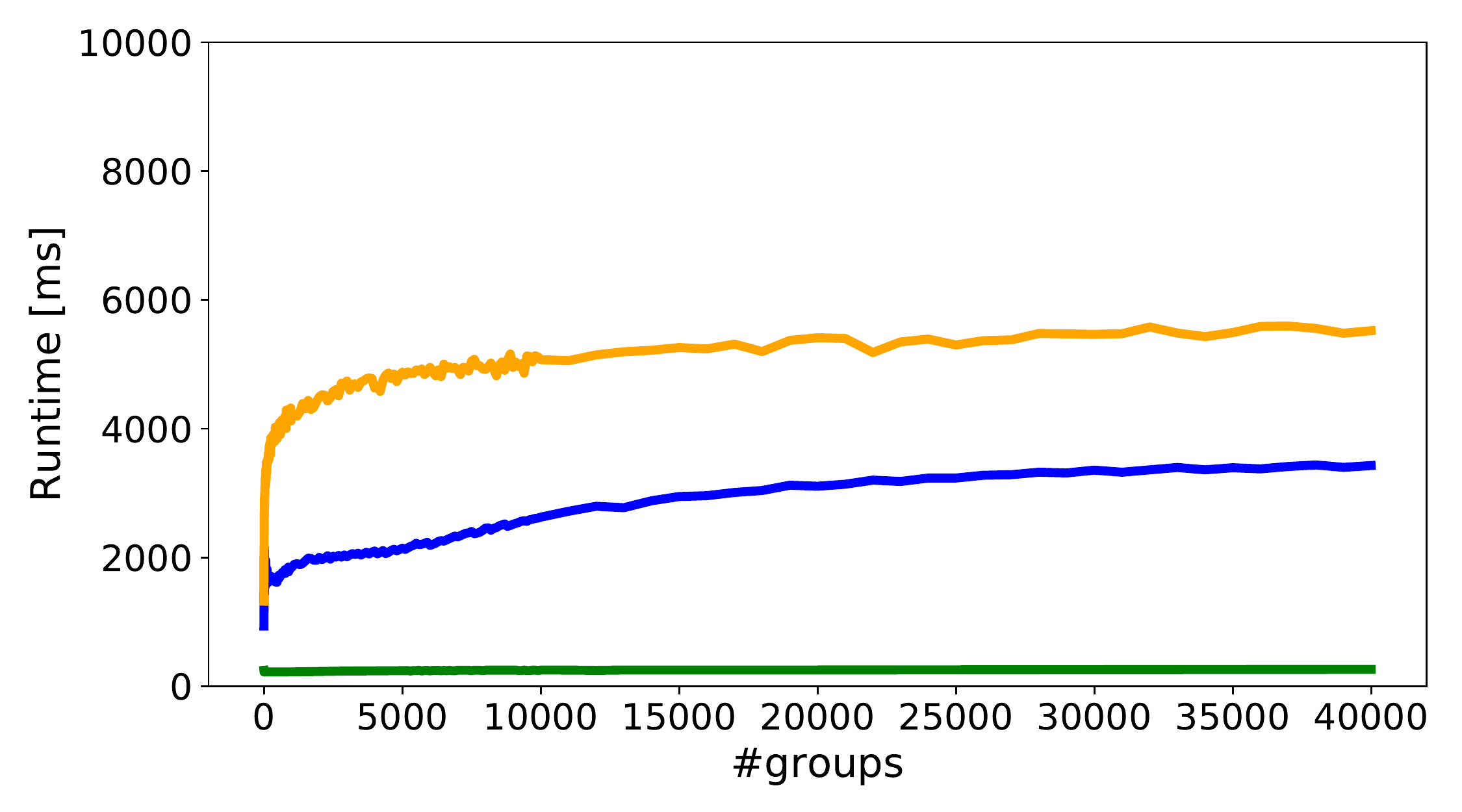}} 
\end{tabular}
\end{center}
\vspace*{-0.2cm}
\caption{\label{fig:grouping}Grouping performance of four different grouping algorithms on four different input datasets.}
\vspace*{0.2cm}
\end{figure}

This section compares the performance of the aforementioned algorithms by their execution time on the four different datasets. Figure~\ref{fig:grouping} shows one plot for each of the the four different datasets. Each plot depicts the execution time in milliseconds for an increasing number of groups.

\noindent\textbf{Sorted \& Dense.} For this combination OG and SPHG exhibit the best performance at roughly 250 ms being more than four times faster than HG. SOG performs even worse because the already sorted input data is again (unnecessary) sorted incurring a considerable overhead. Furthermore, the execution time of each algorithm is mostly independent of the number of groups.

\noindent\textbf{Sorted \& Sparse.} This experiment shows that the execution time of HG, OG, and SOG is essentially independent of the data density. OG again performs best at around 250 ms. Since we have a sparse data domain, we cannot use SPHG. Instead, we use BSG, which incurs logarithmic costs relative to the number of groups. However, no algorithm comes close to the performance of OG.

\noindent\textbf{Unsorted \& Dense.} In this case, SPHG is the best performing algorithm at a constant execution time of roughly 250 ms, being unaffected by the sortedness of the data. The execution time of HG grows with an increasing number of groups because of caching effects. For less than about 500 groups, SOG displays a steep rise in execution time and afterwards merely a modest increase.

\noindent\textbf{Unsorted \& Sparse.} For this setting, we can neither exploit the sortedness nor the density of the data. Without these properties, HG is superior in a wide range of number of groups. However, for up to 14 groups (see zoom-in in the respective plot), BSG outperforms HG. This opens up another optimisation dimension in which the number of distinct values should be considered.  

In summary, this experiment shows that not only sortedness determines the fastest algorithm but also other properties like in this case density.

\subsection{DQO-enabled Dynamic Programming}
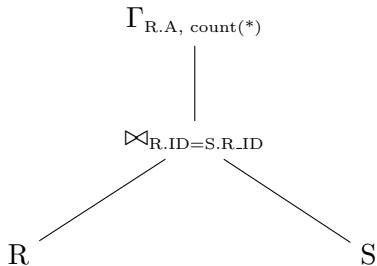
\begin{figure}[h!]
    \centering
    \begin{tikzpicture}[
            op/.style={shape=circle, draw=black, minimum width=2cm}]
        \node (grp) at (0,0) {$\Gamma_{\text{\tiny R.A, count(*)}}$};
        \node (join) [below=of grp] {$\bowtie_{\text{\tiny R.ID=S.R\_ID}}$};
        \node (R) [below left=of join] {R};
        \node (S) [below right=of join] {S};
        % draw lines
        \draw (grp) -- (join);
        \draw (join) -- (R);
        \draw (join) -- (S);
    \end{tikzpicture}
    \vspace*{-0.2cm}
    \caption{\label{fig:log-plan}Logical plan of example query.}
\end{figure}

This section shows how classical dynamic programming can lead to better query execution plans when extending it with DQO.
Consider the following query:
\begin{verbatim}
SELECT R.A, COUNT(*)
FROM R JOIN S ON R.ID=S.R_ID
GROUP BY R.A;
\end{verbatim}
For the physical implementations of the joins, we assume the algorithmic counterparts of our grouping implementations.  The corresponding cost models are shown in Table~\ref{tab:cost-model}.
We assume the size of tables R and S to be $40,000$ and $90,000$, respectively. Further, we assume the output-size of the join to be $90,000$ because of the foreign-key constraint \texttt{R.ID=S.R\_ID} and the output-size to be $20,000$.
\begin{table}[h]
    \begin{scriptsize}
    \centering
    \renewcommand{\arraystretch}{1.15}
    \begin{tabular}{|l||l|l|}
        \hline
                    & \textbf{Grouping}                                  & \textbf{Join}                                                                  \\\hline\hline
        \textbf{hash-based}  & $ HG(R) = 4 \cdot |R|$                 & $ HJ(R,S) = 4 \cdot (|R| + |S|)$                                     \\\hline
        \textbf{order-based} & $ OG(R) = |R| $    & $ OJ(R,S) = |R| + |S|$ \\\hline
        \textbf{sort \& order-based} & $ SOG(R) = |R| \cdot log_2(|R|) + |R|$    & $ SOJ(R,S) = |R| \cdot log_2(|R|) + |S| \cdot log_2(|S|) + |R| + |S|$ \\\hline
        \textbf{static perfect hash-based} & $ SPHG(R) = |R|$                         & $ SPHJ(R,S) = |R| + |S|$                                               \\\hline
        \textbf{binary search-based} & $ BSG(R) = |R| \cdot log_2(\#groups) $   & $ BSJ(R,S) = |R| \cdot log_2(\#groups) + |S| \cdot log_2(\#groups)$   \\\hline
    \end{tabular}
 \vspace*{-0.2cm}
    \caption{\label{tab:cost-model}Cost models for grouping and join-algorithms used.}
    \end{scriptsize}
\end{table}
\indent Figure~\ref{fig:log-plan} shows the logical plan of the above query.

\noindent First, we join both input relations R and S and afterwards, we compute the grouping result on the output of the join.  However, we want to determine which physical implementations of the join and grouping operators lead to the overall best plan.  While SQO only considers data sortedness, DQO also considers other DQO plan properties (cf.~Section~\ref{sec:DQOplanproperties}), here: the density of the grouping keys. This allows us to use static perfect hash-based algorithmic designs (SPH) for our join and grouping implementations.  For such a small query, a classical dynamic programming algorithm is sufficiently fast. In case the search space gets too big, we believe that we can fall back on established dynamic programming variations~\cite{DBLP:conf/vldb/MoerkotteN06, DBLP:conf/sigmod/MoerkotteN08}.

\begin{figure}[h!]
    \centering
    \includegraphics[width=0.7\textwidth]{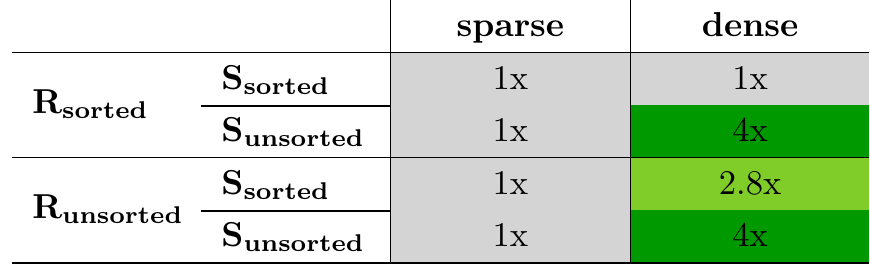}
    \vspace*{-0.2cm}
    \caption{\label{fig:dp-improv-factors}Improvement factors for the estimated plan costs of DQO over SQO.}
\end{figure}

The Table~\ref{tab:dp-pivot} in Appendix~\ref{app:dp-results} shows the DQO dynamic programming tables including the density property, for the different combinations of input relation properties. For each combination, we compared the estimated costs of the best DQO plan to the estimated costs of the best SQO plan. Figure~\ref{fig:dp-improv-factors} shows the improvement factors for the estimated plan costs of DQO over SQO. Compared to DQO, the only difference is that SQO does not consider data density as a meta-relational property. Since SPH can only be used in a dense domain, for sparse data DQO generates the same plans as SQO, resulting in no improvement. In case both inputs are sorted, the order-based implementations achieve the cheapest plans regardless of the data density.  However, if at least one input is unsorted, DQO generates plans with an improvement factor of up to 4x. In this case, DQO chooses plans that use the SPHJ and SPHG algorithms.

In summary, the experiments show that depending on the underlying data properties, different physical implementations lead to the cheapest plans and achieve the best execution time. This supports our claim that more fine-granular optimisation can lead to better query plans.

\section{Related Work}
\label{sec:relatedwork}
Notice, that the bulk of related work has already been discussed inline above or will be discussed in the Research Agenda in Section~\ref{sec:researchagenda}.

Stratos' Idreos work~\cite{DBLP:journals/debu/IdreosZADHKGMQW18, DBLP:conf/sigmod/IdreosZHKG18} is the closest work to ours when it comes to physical `indexing operators', see Section~\ref{sec:researchagenda} (Algorithmic Index Views) for details.

Christoph Koch's work~\cite{DBLP:conf/cidr/Koch13} discusses the problems with missing abstractions in the database community in particular when communicating our findings to the systems community. The Introduction of that paper is a must-read for every researcher: \textit{`... What is frequently lacking is a conclusive deconstruction of the research contribution [of a paper] into first principles and fundamental patterns from which the contribution is composed. ...'} In his work, Christoph Koch develops methods for synthesising and composing algorithms~\cite{DBLP:journals/debu/Koch14}. That work is however much more PL and compiler-oriented than DQO. In particular, we argue that database domain-specific decisions should not be delegated to the PL compiler.

\section{Research Agenda}
\label{sec:researchagenda}
\noindent\textbf{Revisit SQO Algorithms.} We envision that many existing SQO algorithms can easily be extended to support DQO. Just like extending SQO to large queries~\cite{DBLP:conf/sigmod/NeumannR18}, the challenge will be to extend them to deep queries and find the right sweet-spots. Recall that, in the history of SQO, initially only relatively small queries could be optimised, but over time the queries became bigger and bigger. We foresee the same to happen with DQO: over time deeper and deeper queries will become optimisable. As long as optimisation time in DQO is an issue, we need MAVs to the rescue.

\noindent\textbf{Physiological Algebra.} An interesting research subspace will be to identify the right components to use in DQO, i.e.~what are suitable organelles and macro-molecules to consider? We envision that this will lead to a physiological component set akin to relational algebra yet including both logical and physical aspects.

\noindent\textbf{Algorithmic Views Selection.} A promising direction is to systematically research and evaluate MAVs and the Algorithmic Views Selection Problem (AVSP). When to materialise which algorithm into an MAV? Beforehand or at query time? What do we possibly gain or lose at query time? These trade-offs have to be explored carefully. And, yes, for sure: these trade-offs are absolutely workload-dependent. For which parts of a query plan should we consider DQO?

\noindent\textbf{Partial Algorithmic Views.} Rather than fully materialising parts of a deep query plan into an MAV, or, if we pick the other extreme, not materialising it at all, there is an interesting middle-ground: It makes sense to \textit{partially optimise an MAV offline} and leave some flexibility for DQO at query time. Which portions should be left up for DQO at query time? Again, these trade-offs have to be explored carefully. Actually, there are many interesting lessons here that can be adapted from compiler construction in this space.

\noindent\textbf{Materialised Algorithmic Index Views.} An entire interesting research subspace is to apply DQO to indexing.  It would be exciting to explore DQO in that context. In database literature, we witness the birth of about a dozen index structures every year. Most indexes are basically composed of substructures (atoms in our analogy), i.e.~different nodes and leaf types. Class-book index structures like a B-trees and binary search trees use a tiny set of node and leaf types, other indexes extend that set slightly allowing for more heterogeneous trees~\cite{DBLP:conf/icde/LeisK013}.

An extreme version of this is~\cite{DBLP:journals/debu/IdreosZADHKGMQW18,DBLP:conf/sigmod/IdreosZHKG18}. However, a synthesised data structure (SDS) is simply one special case of what we propose. Basically, in DQO, \textit{a synthesised data structure is one particular type of an MAV}. In addition, for an SDS all of the optimisation happens offline. That is an unnecessary restriction as already outlined in Section~\ref{sec:algorithmicviewselection}. In DQO we do not need to synthesise the entire index or any other MAV beforehand. Which implies the following:

\noindent\textbf{Runtime-Adaptivity and Reoptimisation of MAVs.} So far we suggested to optimise deep query plans and \textit{then} execute these plans. As with shallow query plans, the literature on reoptimisation (during query time) as well as adaptivity should be revisited in the light of DQO. 
For instance, in traditional indexing, for each column, the decision whether to create an index is binary. What if we make that decision continuous? Like that different parts of a column are not, slightly, or fully indexed. That is the core idea of adaptive indexing~\cite{DBLP:conf/cidr/KerstenM05,DBLP:journals/pvldb/SchuhknechtJD13}. An adaptive index has built-in heuristics to make these decisions at runtime based on the incoming queries. And even those heuristics may be meta-adapted~\cite{DBLP:conf/icde/SchuhknechtDL18}. In the DQO universe a (meta-)adaptive index is simply a partial MAV where some optimisation decisions have been delegated to query time and baked into that MAV. This idea should be revisited for all physical components currently used in SQO; not only indexes. 

\noindent\textbf{Longterm Vision.} We are planning to integrate DQO into mutable~\cite{mutable}. Mutable is an extensible research database system built at Saarland University. In particular, we want to explore how to make a smooth transition from SQO to DQO and find the sweet-spots for any given workload.

\section{Conclusions}
This paper made several contributions: we opened the book for Deep Query Optimisation (DQO). We presented the general
idea, contrasted it to SQO, and showed the high potential of DQO. In addition, we introduced the concept of Materialised Algorithmic Views and the Algorithmic View Selection Problem. We presented early experimental results with DQO. In addition, we compiled a research agenda.

\bibliographystyle{abbrv}
\bibliography{references}

\appendix
\section{Dynamic Programming Results}
\label{app:dp-results}
Table~\ref{tab:dp-pivot} shows the dynamic programming tables for the different combinations of input relation properties in a DQO setting, i.e.~we consider additional meta-relational properties. Here, we include the density property in addition to the sortedness of a relation. We consider static perfect hash-based algorithms only for dense data and binary search-based algorithms only for sparse data. In addition, we assume sort and order-based algorithms to incur sorting cost only if the input is not sorted.  Furthermore, if at least one of the inputs is not sorted, we do not consider order-based algorithms.

The dynamic programming results show that in a dense domain, the usage of specialised algorithms like SPH leads to plans with the lowest estimated cost.

\begin{table}[!t]
    \centering
    \resizebox{\textwidth}{!}{
    \begin{tabular}{%
            >{\raggedright\arraybackslash}p{0.4em}%
            >{\raggedright\arraybackslash}p{0.4em}%
            >{\raggedright\arraybackslash}p{0.4em}%
            |c|c}%
                            &                                           &                           & \multicolumn{2}{c}{\textbf{S}}     \\           
                            &                                           &                           & \textbf{sorted} & \textbf{unsorted}  \\\hline
        \multirow{4}{*}[-6.2em]{\hspace{-0.4em}\textbf{R}}  & \multirow{2}{*}{\rotatebox{90}{\textbf{dense}}}    &
        \multirow{1}{*}[4.5em]{\rotatebox{90}{\textbf{sorted}}}    & \includegraphics{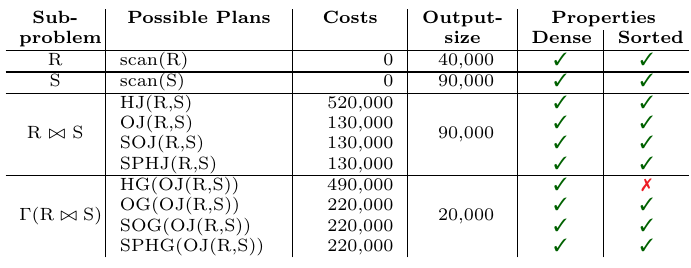} & \includegraphics{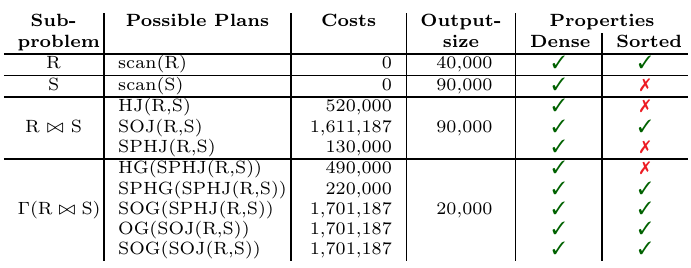} \\\cline{3-5}
                            &                                           &
                            \multirow{1}{*}[5.5em]{\rotatebox{90}{\textbf{unsorted}}}  & \includegraphics{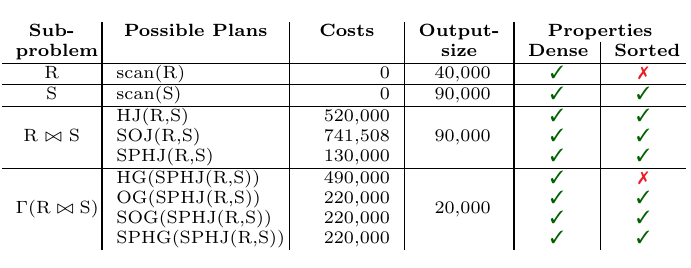} & \includegraphics{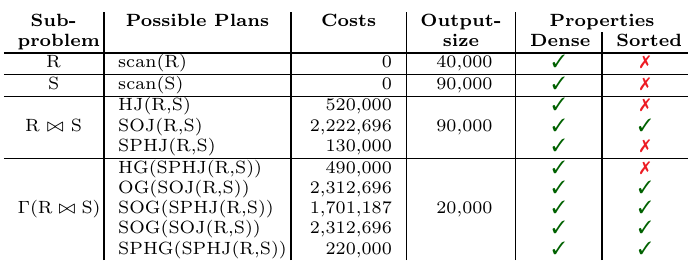} \\\cline{2-5}
                            & \multirow{2}{*}{\rotatebox{90}{\textbf{sparse}}}   & \multirow{1}{*}[4.5em]{\rotatebox{90}{\textbf{sorted}}}    & \includegraphics{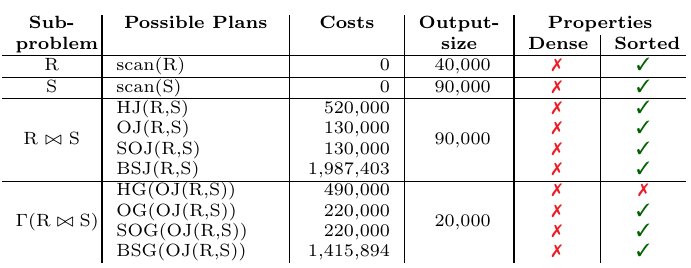} & \includegraphics{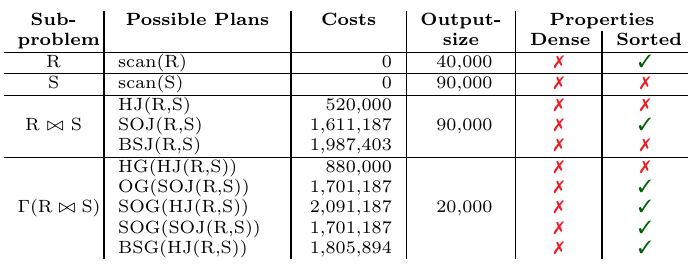} \\\cline{3-5}
                            &                                           & \multirow{1}{*}[5.5em]{\rotatebox{90}{\textbf{unsorted}}}  & \includegraphics{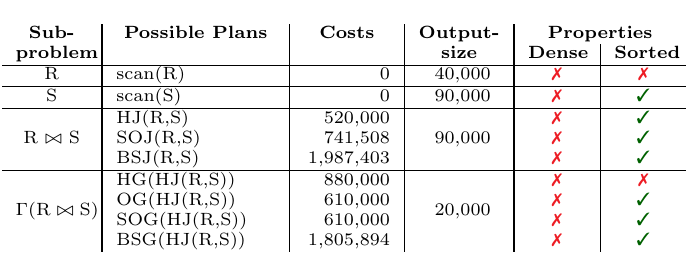} & \includegraphics{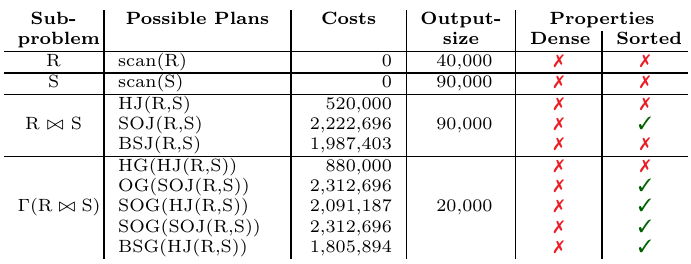} \\\hline
    \end{tabular}}
    \vspace*{-0.2cm}
    \caption{\label{tab:dp-pivot}Pivot table of dynamic programming tables for different combinations of input relation properties. Notice that we assume that density/sparseness is the same on the key and foreign key columns on both input relations.}
\end{table}

\end{document}